\documentclass[aps,a4paper,superscriptaddress,nofootinbib,twocolumn]{revtex4-1}
\usepackage{amsmath,amssymb,gensymb,graphicx,multirow,dcolumn,bm,latexsym,soul,nicefrac,acronym}
\usepackage[mathscr]{euscript}
\usepackage{appendix}
\usepackage[colorlinks,linkcolor=blue,citecolor=blue,urlcolor=blue ]{hyperref}
\usepackage[normalem]{ulem}
\usepackage{float}
\usepackage{soul}
\usepackage{verbatim}
\usepackage{color}
\usepackage{acronym}
\usepackage{subfigure}
\usepackage{appendix}
\usepackage{longtable}
\usepackage{footnote}
\usepackage{cancel}
\usepackage{ulem} 

\newcommand{\eq}[1]{Eq.~(\ref{#1})}
\newcommand{\fig}[1]{Fig.~\ref{#1}}

\newcommand{\TRC}{MOE Key Laboratory of TianQin Mission, TianQin Research Center for Gravitational Physics $\&$  School of Physics and Astronomy, Frontiers Science Center for TianQin, CNSA Research Center for Gravitational Waves, Sun Yat-sen University (Zhuhai Campus), Zhuhai 519082, China}

\begin{document}

\title{Mapping Anisotropies in the Stochastic Gravitational-Wave Background with TianQin}
\author{Zhi-Yuan Li}
\affiliation{\TRC}
%\affiliation{\HUST}
\author{Zheng-Cheng Liang}
\email{Corresponding author: liangzhch7@mail.sysu.edu.cn}
\affiliation{\TRC}
\author{En-Kun Li}
\affiliation{\TRC}
\author{Jian-dong Zhang}
\affiliation{\TRC}
\author{Yi-Ming Hu}
\email{Corresponding author: huyiming@sysu.edu.cn}
\affiliation{\TRC}

\date{\today}

\begin{abstract}
% Importance/motivation
In the milli-Hertz frequency band, stochastic gravitational-wave background can be composed of both astronomical and cosmological sources, both can be anisotropic. 
Numerically depicting these anisotropies can be critical in revealing the underlying properties of their origins. 
% what we did (for the first time)
For the first time, we perform a theoretical analysis of the constraining ability of TianQin on multiple moments of the stochastic background. 
% what we found: 1. l
First, we find that with a one-year operation, for a background with a signal-to-noise ratio of 16, TianQin can recover the multiple moments up to $l=4$.
% 2. mirror symmetry
We also identified a unique feature of the stochastic background sky map, which is the mirror symmetry along the fixed orbital plane of TianQin.
% 3. threshold 
Thirdly, we explain the difference in anisotropy recovering ability between TianQin and LISA, by employing the criteria of the singularity of the covariance matrix (which is the condition number).
% 4. AET/XYZ
Finally, we find that since the different data channel combinations correspond to different singularities, certain combinations might have an advantage in stochastic background map-making.
% usage of this work
We believe that the findings of this work can provide an important reference to future stochastic background analysis pipelines.
It can also serve as a guideline for designing better gravitational-wave detectors aiming to decipher anisotropies in the stochastic background. 

%
%In this work, we examine the capability of TianQin to map the anisortropy of stochastic gravitational-wave background across various multipole moments. 
%We conduct map-making based on a maximum likelihood method by correlating data of time-delay interferometry channel sets. 
%Despite noise in the data, TianQin can effectively identify multipole moments of order $l$ up to 4 for a point source within 1 year of operation, achieving a signal-to-noise ratio of around 16. 
%A notable limitation stems from the fixed pointing direction of TianQin, introducing a plane symmetry into the map-making process. 
%We further explore other factors influencing this process, revealing that stochastic backgrounds with larger spectral indices yield more precise map-making results. 
%Additionally, the well-considered selection of channel sets for data analysis can significantly enhance the accuracy of map-making. 
%These insights can be not only pertinent to TianQin but also extend to other space-borne detectors, such as the Laser Interferometer Space Antenna. 
\end{abstract}

\maketitle
\acrodef{SGWB}{stochastic \ac{GW} background}
\acrodef{GW}{gravitational-wave}
\acrodef{CBC}{compact binary coalescence}
\acrodef{MBHB}{supermassive black hole binary}
\acrodef{BBH}{binary black hole}
\acrodef{BNS}{binary neutron star}
\acrodef{EMRI}{extreme-mass-ratio inspiral}
\acrodef{DWD}{double white dwarf}
\acrodef{GDWD}{Galactic double white dwarf}
\acrodef{EDWD}{extragalactic double white dwarf}
\acrodef{BH}{black hole}
\acrodef{NS}{neutron star}
\acrodef{BNS}{binary neutron star}
\acrodef{LIGO}{Laser Interferometer Gravitational-Wave Observatory}
\acrodef{LISA}{Laser Interferometer Space Antenna}
\acrodef{TQ}{TianQin}
\acrodef{KAGRA}{Kamioka Gravitational Wave Detector}
\acrodef{ET}{Einstein telescope}
\acrodef{DECIGO}{DECi-hertz interferometry GravitationalWave Observatory}
\acrodef{CE}{Cosmic Explorer}
\acrodef{NANOGrav}{The North American Nanohertz Observatory for Gravitational Waves}
\acrodef{LHS}{left-hand side}
\acrodef{RHS}{right-hand side}
\acrodef{ORF}{overlap reduction function}
\acrodef{ASD}{amplitude spectral density}
\acrodef{PSD}{power spectral density}
\acrodef{SNR}{signal-to-noise ratio}
\acrodef{FIM}{Fisher information matrix}
\acrodef{TDI}{time delay interferometry}
\acrodef{PIS}{peak-integrated sensitivity}
\acrodef{PT}{phase transitions}
\acrodef{PLIS}{power-law integrated sensitivity}
\acrodef{GR}{general relativity}
\acrodef{PBH}{primordial black hole}
\acrodef{SSB}{solar system baryo}
\acrodef{PTA}{Pulsar Timing Arrays}
\acrodef{SVD}{singular-value decomposition}
\acrodef{JSD}{Jensen-Shannon divergence}

\section{Introduction}
A \ac{SGWB} consists of \acp{GW} that, when individually undetectable, collectively form a background~\cite{Romano:2016dpx}.~Various mechanisms contribute to the \ac{SGWB}, which can be categorized into astrophysical and cosmological origins~\cite{Christensen:2018iqi,Romano:2019yrj,LISACosmologyWorkingGroup:2022kbp}. 
Astrophysical origins primarily include \acp{DWD}~\cite{Huang:2020rjf,Korol:2017qcx}, massive \acp{BBH}~\cite{Wang:2019ryf}, stellar-mass \acp{BBH}~\cite{Liu_2020,Wang:2023tle}, \acp{BNS} ~\cite{Surace:2015ppq,Talukder:2014eba}, extreme-mass-ratio inspirals (EMRIs)~\cite{Fan:2020zhy,Zi:2021pdp,Ye:2023uvh}, and \ac{GW} bursts~\cite{Wu:2023rpn}. 
Cosmological origins mainly contain inflation~\cite{Guth:1982ec}, first-order \acp{PT}~\cite{Hogan:1983ixn}, and cosmic defects~\cite{Kibble:1976sj,LISACosmologyWorkingGroup:2022jok}. 
\ac{SGWB} from either astrophysical origin~\cite{Wang:2019ryf,Huang:2020rjf,Liu:2020eko,Fan:2020zhy} or cosmological origin~\cite{Siemens:2006yp,Figueroa:2012kw,Mandic:2016lcn,Saikawa:2018rcs,Caprini:2018mtu,Auclair:2019wcv,Wang:2020jrd} can exhibit the potential for anisotropy~\cite{Allen:1996gp}. 
The anisotropy can offer insights into the distribution of compact binaries within our Galaxy, the history of galaxy mergers~\cite{Cusin:2018rsq,Capurri:2021zli,Bellomo:2021mer}, as well as the priceless primordial information of the early Universe~\cite{Geller:2018mwu,Jenkins:2018nty,Li:2021iva,Wang:2021djr,Profumo:2023ybp,Schulze:2023ich}. 
For example, the structure of our Galaxy can be mapped using the lower multipole moments with $l\le 4$~\cite{Ungarelli:2001xu,Breivik:2019oar}. 
Conversely, the higher multipole moments, with $l\gtrsim 100$, are more suitable for detecting the subtle imprints of primordial fluctuations that are preserved in cosmological backgrounds~\cite{Li:2021iva,LISACosmologyWorkingGroup:2022kbp}.

The diverse origins of the \ac{SGWB} allow it to span multiple frequency bands. 
At frequencies in the hundreds of hertz, ground-based detectors, despite not yet detecting \acp{SGWB}, have set an upper limit on the dimensionless energy density~\cite{KAGRA:2021kbb}. 
In the nano-hertz (nHz) frequency band, recent findings from \acp{PTA} have provided compelling evidence for the existence of \ac{SGWB}~\cite{NANOGrav:2023gor,Xu:2023wog,EPTA:2023fyk,Reardon:2023gzh}. 
Moving to the milli-hertz (mHz) frequency band, space-borne \ac{GW} detectors, such as TianQin~\cite{TianQin:2015yph} and \ac{LISA}~\cite{LISA:2017pwj}, are expected to detect a Galactic foreground~\cite{Benacquista:2005tm,Liang:2021bde}.

When conducting \ac{GW} detection, space-borne detectors face the challenge of canceling laser noise due to their motion. 
This issue can be addressed through \ac{TDI} techniques, which enable the formation of various types of TDI channels~\cite{Tinto:1999yr,Shaddock:2003bc,Vallisneri:2005ji,Muratore:2020mdf,Tinto:2020fcc}. 
Among all the \ac{TDI} channels, the unequal-arm Michelson, typically represented by the XYZ channel set, is the most widely utilized. 
This set can be further recombined to form the orthogonal AET channel set, with the T channel specifically designed to be insensitive to \acp{GW}. 
By employing the T channel as a noise monitor, the \ac{SGWB} signal can be extracted through the correlation of data from other channels, which is often referred to as the null-channel method~\cite{Tinto:2001ii,Hogan:2001jn,Adams:2010vc,Smith:2019wny,Boileau:2020rpg,Muratore:2021uqj,Cheng:2022vct,Gowling:2022pzb,Alvey:2023npw}.

From the \ac{SGWB} signal, one can derive a distribution of \ac{GW} power across the sky, know as a sky map, to uncover the anisotropy of the \ac{SGWB}. 
This process is termed ``map-making". 
Initially, the \ac{SGWB} signal produces a dirty map, which reflects the \ac{GW} sky map as observed by detectors. 
The dirty map is susceptible to alterations due to variations in the detector's beam pattern. 
To refine the dirty map into a pure \ac{GW} sky map, or clean map, it is essential to solve the deconvolution problem~\cite{Cornish:2001hg,Taruya:2005yf,Taruya:2006kqa}. 
Thrane et al. are among the first to tackle this challenge~\cite{Thrane:2009fp}, employing maximum likelihood estimates for the dirty map and then deconvolving it to recover a clean map~\cite{Renzini:2018vkx,Suresh:2020khz,Agarwal:2021gvz,Renzini:2021iim,Xiao:2022uvq}. 
Building on this foundation, Renzini et al. have introduced a frequentist maximum likelihood method to map \ac{GW} power~\cite{Contaldi:2020rht}, conducting an all-sky analysis based on the complete likelihood of the data. 
In addition to the maximum likelihood method, a Bayesian spherical harmonic method has recently been proposed~\cite{Banagiri:2021ovv,Renzini:2022alw,Rieck:2023pej}.

In this paper, we harness maximum likelihood estimation to perform map-making with TianQin. 
In order to quantify the ability of TianQin, we simulate a Gaussian, stationary, and unpolarized point source into the data, both in the absence and presence of noise. 
To furnish a thorough analysis of the map-making process, we also examine the influence of several key factors, including the detector design, the spectral characteristics of the \ac{SGWB}, the selected frequency range, and the specific \ac{TDI} channel sets employed for the data analysis.

The structure of the paper is organized as follows: we review the formalism of the \ac{SGWB} detection in Sec.~\ref{sec:basic}. 
The maximum likelihood method and the deconvolution process for map-making are detailed in~Sec.~\ref{sec:method}. 
Based on the knowledge from the previous sections, we present the set up and results of map-making in~\ref{sec:result}. 
Finally, we give a brief summary in Sec.~\ref{sec:summary}.

\section{Formalism}\label{sec:basic}
\subsection{Energy spectral density of stochastic background}\label{Energy}
An \ac{SGWB} is composed of a large number of \acp{GW}. 
Consequently, in transverse-traceless coordinates, the metric perturbations $h(t,\vec{x})$ associated with the \ac{SGWB} can be represented as a superposition of sinusoidal plane waves, characterized by frequency $f$ and the unit vector $\hat{\mathbf{n}}$ indicating the line-of-sight direction on the sky:
\begin{equation}
\begin{split}
    h_{ij}(t,\vec{x}) = &\sum_{P = +, \times} 
    \int ^{\infty}_{-\infty}{\rm d}f 
    \int_{S^2} {\rm d}\hat{\Omega}_{\hat{\mathbf{n}}}\,
    e_{ij}^P(\hat{\mathbf{n}})\widetilde{h}_P(f,\hat{\mathbf{n}}) \\
    &\times {\rm e}^{{\rm i} 2\pi f [ t+\hat{\mathbf{n}}\cdot \vec{x}(t)/c ]},
\end{split}
\end{equation}
where $\vec{x}$ denotes the location at which the \ac{GW} measurement is conducted, $c$ is the speed of light. 
The symbol $e_{ij}^P(\hat{\mathbf{n}})$ signifies the \ac{GW} polarization tensor corresponding to the polarization $P$, while $\widetilde{h}_P(f,\hat{\mathbf{n}})$ denotes the Fourier amplitude. 

In this work, we focus on the anisotropic, Gaussian, stationary, and unpolarized \ac{SGWB}. 
Thus, the Fourier amplitude is treated as a zero-mean random variable, characterized by the one-sided \ac{PSD} $\mathscr{P}_{\rm h}$:
\begin{equation}
	\label{eq:hh}
    \langle \widetilde{h}_P(f,\hat{\mathbf{n}})\widetilde{h}^{\ast}_{P'}(f,\hat{\mathbf{n}}') \rangle = \frac{1}{4}\delta(f-f')\delta_{PP'}\delta^2(\hat{\mathbf{n}}-\hat{\mathbf{n}}')\mathscr{P}_{\rm h}(f,\hat{\mathbf{n}}),
\end{equation}
where the symbol * denotes the conjugate. 
The factor of 1/4 pertains to the definition of one-sided \ac{PSD}, encompassing the summation over both polarizations. 
Given the assumption that the direction and intensity of the \ac{SGWB} are independent, the $\mathscr{P}_{\rm h}(f,\hat{\mathbf{n}})$ can be factorized into the spectral shape $\bar{H}(f)$ and angular distribution $\mathcal{P}_{\rm h}(\hat{\mathbf{n}})$:
\begin{equation}
	\label{eq:Ph}
    \mathscr{P}_{\rm h}(f,\hat{\mathbf{n}}) = \bar{H}(f)\mathcal{P}_{\rm h}(\hat{\mathbf{n}}).
\end{equation}
The angular distribution $\mathcal{P}_{\rm h}(\hat{\mathbf{n}})$ can be expanded in terms of a set of basis:
\begin{equation}
    \mathcal{P}_{\rm h}(\hat{\mathbf{n}}) = \sum_{\alpha} \mathcal{P}_{\alpha}e_{\alpha}(\hat{\mathbf{n}}),
\end{equation}
where $e_{\alpha}(\hat{\mathbf{n}})$ can be spherical harmonic or pixel basis. 
%is the arbitrary basis functions, for a diffuse background, dominated by a dipole or quadrupolar distribution, a spherical harmonic decomposition is a good choice:
%\begin{equation}\label{eq:spherical}
    %\mathcal{P}_{\rm h}(\hat{\mathbf{n}}) = \frac{1}{\sqrt{4\pi \,P_{00}}}\sum_{l=0}^{\infty}\sum_{m=-l}^{l}P_{lm}Y_{lm}(\hat{\mathbf{n}}),
%\end{equation}
%where $P_{lm}$ are the spherical harmonic coefficients of the signal and $Y_{lm}(\hat{\mathbf{n}})$ are the spherical harmonic basis functions.
%$1/\sqrt{4\pi\,P_{00}}$ is the normalization factor. 
%In this paper, we choose the pixel basis: 
%\begin{equation}\label{eq:pixel}
    %\mathcal{P}_{\rm h}(\hat{\mathbf{n}}) = \mathcal{P}_{\rm \hat{\mathbf{n}}}\delta(\hat{\mathbf{n}},\hat{\mathbf{n}}').
%\end{equation}
%Here, $\mathcal{P}_{\rm \hat{\mathbf{n}}}$ specifies the power contained in each pixel. %Both~\eq{eq:spherical} and~\eq{eq:pixel} guarantee that
%\begin{equation}
	%\label{normalize direction}
    %\int_{S^2}{\rm d}\hat{\Omega}_{\hat{\mathbf{n}}}\,
    %\mathcal{P}_{\rm h}(\hat{\mathbf{n}}) = 1.
%\end{equation}
With the definitions provided above, the dimensionless energy density spectrum $\Omega_{\rm gw}$ can be derived as follows~\cite{Thrane:2009fp}:
\begin{equation}
	\label{eq:O2P}
	\Omega_{\rm gw}(f)
	=
	\frac{1}{\rho_c}\frac{\rm{d}\rho_{\rm gw}}{{\rm d}(\ln f)}
	=
	\frac{2\pi^2 f^3}{3H^2_0}
	\int_{S^2} {\rm d}\hat{\Omega}_{\hat{\mathbf{n}}}\,
	\mathscr{P}_{\rm h}(f,\hat{\mathbf{n}}),
\end{equation}
where the critical energy density $\rho_c=3H^2_0c^2/{8\pi} G$, incorporating the gravitational constant $G$ and the Hubble constant $H_0$. 
The term ${\rm d}\rho_{\rm gw}$ represents the \ac{GW} energy density contained within the frequency interval [$f$,$f+{\rm d}f$].
Generally speaking, the energy density spectrum can be characterized by a power-law form with an index $\alpha$ and a reference frequency $f_{\rm r}$:
\begin{equation}
	\label{eq:Omega}
    \Omega_{\rm gw}(f) 
    = \Omega_{\rm r}\left(\frac{f}{f_{\rm r}}\right)^{\alpha},
\end{equation}
where $\Omega_{\rm r}$ denotes the reference energy density.

Thus far, we has focused solely on the intrinsic anisotropy of the \ac{SGWB}. 
However, the \ac{SGWB} can also exhibit a degree of kinematic Doppler anisotropy, analogous to observations made in the cosmic microwave background (CMB)~\cite{Kosowsky:2010jm,Cusin:2022cbb,Cruz:2024svc,Heisenberg:2024var}. 
The Doppler effect can cause shifts in both the frequency and direction of the \ac{SGWB}. 
By defining the unit vector $\hat{\mathbf{v}}$ to indicate the direction of relative motion between the \ac{SGWB} rest frame and the moving frame, and $\hat{\mathbf{n}}_{\rm D}$ to signify the direction of the \ac{SGWB} within the moving frame, the coefficient of frequency change due to the Doppler effect can be determined by
\begin{equation}
	\label{eq:dpl_D}
	\mathcal{D} = \frac{\sqrt{1-\beta^2}}{1-\beta\, \xi},
\end{equation}
and the direction shift can be expressed as
\begin{equation}
	\label{eq:dpl_n}
	\hat{\mathbf{n}} = \frac{\hat{\mathbf{n}}_{\rm D}+\hat{\mathbf{v}}\left[(\gamma-1)\xi-\gamma\beta \right]}{\gamma(1-\beta\xi)}.
\end{equation}
Here, $\beta = |\vec{\mathbf{v}}|/c$, $\xi=\hat{\mathbf{n}}_{\rm D}\cdot \hat{\mathbf{v}}$. 
Based on these definitions, the transformation of the energy density between the two frames can be written as:
\begin{equation}
	\label{eq:Omega_D}
    \Omega^{\rm D}_{\rm GW}(f) = 
    \int_{S^2} {\rm d}\hat{\Omega}_{\hat{\mathbf{n}}}\,
    \mathcal{D}^4 \Omega_{\rm GW}\left(\mathcal{D}^{-1}f, \frac{\hat{\mathbf{n}}+\hat{\mathbf{v}}\left[(\gamma-1)\xi-\gamma\beta \right]}{\gamma(1-\beta\xi)}\right).
\end{equation}
Further details regarding this topic can be found in Ref.~\cite{Cusin:2022cbb}. 

\subsection{Detector noise}\label{noise}
Among the various space missions proposed for \ac{GW} detection, this paper primarily focuses on TianQin~\cite{TianQin:2015yph} and \ac{LISA}~\cite{LISA:2017pwj}.  
TianQin will consist of three identical drag-free satellites arranged in an equilateral triangular constellation orbiting the Earth. 
The orbital plane of TianQin is designed to always direct toward to J0806 $(\theta_s = -4.7^{\circ},\phi_s = 120.5^{\circ})$, with each satellite separated approximately $1.7\times 10^5\,\,\rm km$. 
Operating on a ``three months on + three months off" mode, TianQin will have an operational duration of six months each year. 
\ac{LISA}, on the other hand, is designed to orbit the Sun, trailing Earth by about $20\degree$, with three satellites separated about $2.5\times10^{6}\,\,{\rm km}$ apart. 
For the sake of convenience, shorthand notations will be employed in figures and equations: TQ for TianQin, and LS for LISA.

Given the motion of space-borne detectors, phase noise becomes the dominant noise in \ac{GW} detection. 
Fortunately, the phase noise can be depressed by several orders of magnitude through the implementation of \ac{TDI} channels~\cite{Tinto:2022zmf,Tinto:2002de,Wang:2022nea,Muratore:2021uqj,Otto:2015erp}. 
Despite such alleviation, other types of noise, including position noise and acceleration noise, persist as significant considerations. 
For the conventional \ac{TDI} channel set XYZ, which utilizes each satellite in harmony with its neighboring arm, the auto- and cross-\ac{PSD} manifest as follows~\cite{Prince:2002hp,Liang:2022ufy}:
\begin{equation}
\begin{split}
	\label{eq:N_xyz}
    N_{II} &= \frac{4}{L^2}\sin ^2 \left[ \frac{f}{f_{\ast}} \right] \left[S_{\rm p}(f)+2\left(\cos^2\left[\frac{f}{f_{\ast}}\right]+1\right)\frac{S_{\rm a}(f)}{(2\pi f)^4}\right]\\
    N_{IJ} &= -\frac{2}{L^2}\sin^2\left[\frac{f}{f_{\ast}}\right] \cos{\left[ \frac{f}{f_{\ast}} \right]}\left(S_{\rm p}(f)+\frac{4S_{\rm a}(f)}{(2\pi f)^4}\right),
\end{split}
\end{equation}
where $I$, $J$ label the X/Y/Z channels, with $I\neq J$. 
$S_{\rm p}$ and $S_{\rm a}$ represent the position noise and the acceleration noise, respectively. 
Given the arm length $L$, the characteristic frequency $f_{\ast} = c/2\pi L$. 
For TianQin and \ac{LISA}, the characteristic frequencies $f^{\rm TQ}_{*}$ and $f^{\rm LS}_{*}$ are estimated to be approximately 0.28 Hz and 0.019 Hz. 

Based on the channel set XYZ, one can further construct channel set AET by
\begin{equation}
	\begin{split}
	\label{eq:AET}
	{\rm A}&=\frac{1}{\sqrt{2}}({\rm Z}-{\rm X}),\\
	{\rm E}&=\frac{1}{\sqrt{6}}({\rm X}-2{\rm Y}+{\rm Z}),\\
	{\rm T}&=\frac{1}{\sqrt{3}}({\rm X}+{\rm Y}+{\rm Z}).
\end{split}
\end{equation}
With the above construction, the auto-\ac{PSD} of channel set AET can be derived from the the auto- and cross-\ac{PSD} of channel set XYZ, as shown in~\eq{eq:N_xyz}~\cite{Liang:2022ufy}. 
Furthermore, the cross-\ac{PSD} of channel set AET is identically zero, thereby ensuring that the three channels are noise-orthogonal.

\subsection{Detector response}
In addition to detector noise, detector response is another important aspect to \ac{GW} detection. 
The detector motion can lead to variations in the response to \acp{GW} over time. Hence, a long data stream needs to be divided into numerous segments. 
These segments should be long enough to cover the sensitive bands of the detector after transformation from the time domain to the frequency domain. 
Meanwhile, they should also be short enough to ensure that the response within each segment can be regarded as stationary, allowing the application of a short-term Fourier transform~\cite{Romano:2016dpx}.

Within a time interval $\tau$ wherein the response of channel $I$ can be considered stationary, the frequency domain \ac{SGWB} signal
\begin{equation}
	\label{eq:hf}
	\widetilde{h}_{I}(f,\tau)
	=\sum_{P}\int_{S^{2}}\,{\rm{d}}\hat{\Omega}_{\hat{k}}\,
	F_{I}^{P}(f,\tau,\hat{\mathbf{n}})\widetilde{h}_{P}(f,\hat{\mathbf{n}}) 
	\,{\rm e}^{{\rm i}2\pi f\hat{\mathbf{n}}\cdot\vec{x}(\tau)/c},
\end{equation}
where $F_{I}^{P}$ denotes the response function. For the channel set XYZ,
 
\begin{equation}
\begin{aligned}
	F_{I}^P(f,\tau,\hat{\mathbf{n}}) =&2\sin^{2}\left(\frac{f}{f_{\ast}}\right)
	\bigg[\hat{u}^a(\tau) \hat{u}^b(\tau) \mathcal{T}\left(f,\hat{\mathbf{n}}, \hat{u}(\tau)\right) \\
	&-\hat{v}^a(\tau) \hat{v}^b(\tau) \mathcal{T}(f, \hat{\mathbf{n}}, \hat{v}(\tau))\bigg]e^P_{ab}(\hat{\mathbf{n}}),
\end{aligned}
\end{equation}
where $\hat{u}$ and $\hat{v}$ are the orientations of arms associated with the channel $I$. 
The transfer function
\begin{equation}
	\label{eq:T_fn}
	\begin{aligned}
		\mathcal{T}[f,\hat{u}(\tau),\hat{\mathbf{n}}]= & \frac{1}{2}\left[\operatorname{sinc}\left[\frac{f}{2 f_*}(1+\hat{\mathbf{n}} \cdot \hat{u}(\tau))\right] e^{-i \frac{f}{2 f_*}[3-\hat{\mathbf{n}} \cdot \hat{u}(\tau)]}\right. \\
		& \left.+\operatorname{sinc}\left[\frac{f}{2 f_*}(1-\hat{\mathbf{n}} \cdot \hat{u}(\tau))\right] e^{-i \frac{f}{2 f_*}[1-\hat{\mathbf{n}} \cdot \hat{u}(\tau)]}\right], 
	\end{aligned}
\end{equation}
with $\operatorname{sinc} (x) = \sin x/x$.

Given the rotational symmetry of the X/Y/Z channels, an inherent equivalence exists among them in the \ac{SGWB} detection. 
However, the A/E/T channels exhibit different responses: the A/E channels are responsive to \acp{GW}, whereas the T channel is typically less responsive unless it captures high-frequency \acp{GW}. 
Consequently, the T channel serves as a tool for monitoring detector noise~\cite{Hogan:2001jn,Adams:2010vc}. 
Furthermore, the response for A/E/T channels is not merely a linear superposition of the response for X/Y/Z channels; it also contains some exponential terms that significantly impact the response for the T channel. 
The specific form can be found in Eq. (18) of Ref.~\cite{Liang:2022ufy}.

By incorporating the response function and the varying arrival times of the signal at each channel, one can further define the antenna pattern~\cite{Liang:2023fdf}:
\begin{equation}\label{eq: antenna pattern}
	\gamma_{IJ}(f,\tau,\hat{\mathbf{n}})=
	\frac{1}{2} 
	\sum_{P} F_I^P(f,\tau,\hat{\mathbf{n}}) F_J^{P*}(f,\tau,\hat{\mathbf{n}})\,
	{\rm e}^{{\rm i} 2 \pi f \hat{\mathbf{n}} \Delta\vec{x}/c},
\end{equation}
where the factor of $1/2$ arises from the average of polarization. 
The separation vector between channels $\Delta \vec{x}=\vec{x}_{I}-\vec{x}_{J}$. 
Notably, when $I=J$, the antenna pattern refers to the auto-correlation of individual channels. 
Integrating the antenna pattern over all sky yields the \ac{ORF}, which quantifies the reduction in correlation due to the non-parallel alignment and time delay between channels:
\begin{equation}
	\Gamma_{IJ}(f,\tau)= \frac{1}{4\pi}\int_{S^2}d^2\hat{\Omega}_{\hat{\mathbf{n}}}\gamma_{I J}(f,\tau,\hat{\mathbf{n}}),
\end{equation}
where the factor of $1/4\pi$ ensures that the \ac{ORF} is properly scaled.

\begin{figure}
	\centering
	\includegraphics[width=0.5\textwidth]{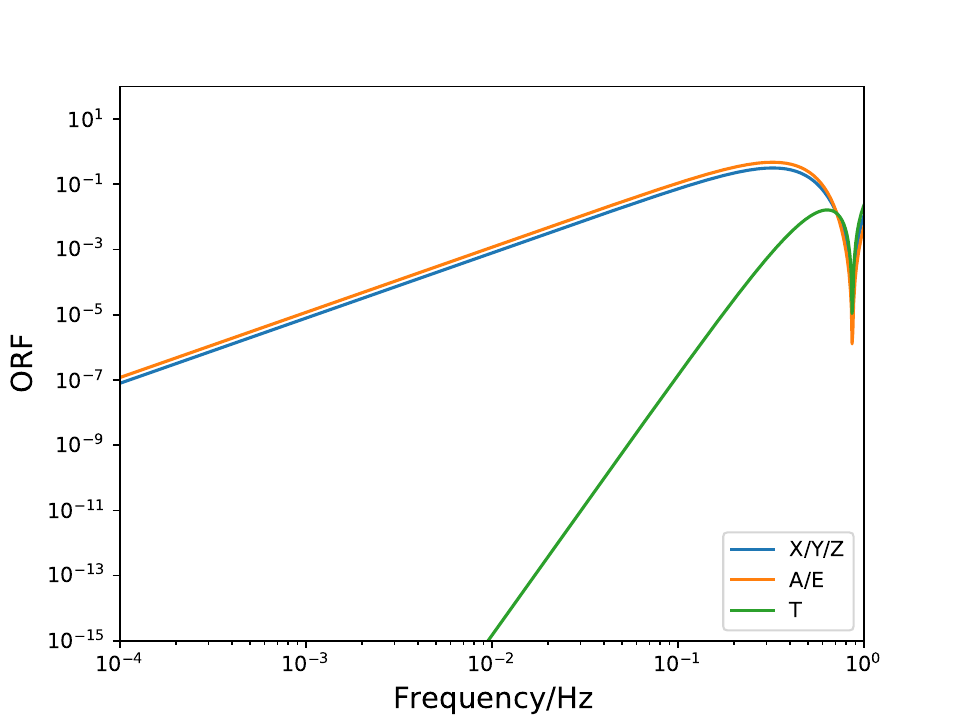}
	\caption{\ac{ORF} of single channels for TianQin. The blue, orange, and green lines are denote the X/Y/Z channels, the A/E channels, and the T channel, respectively.}
	\label{TQ_AT_ORF}
\end{figure}

\begin{figure*}
	\centering
	\includegraphics[width=0.85\textwidth]{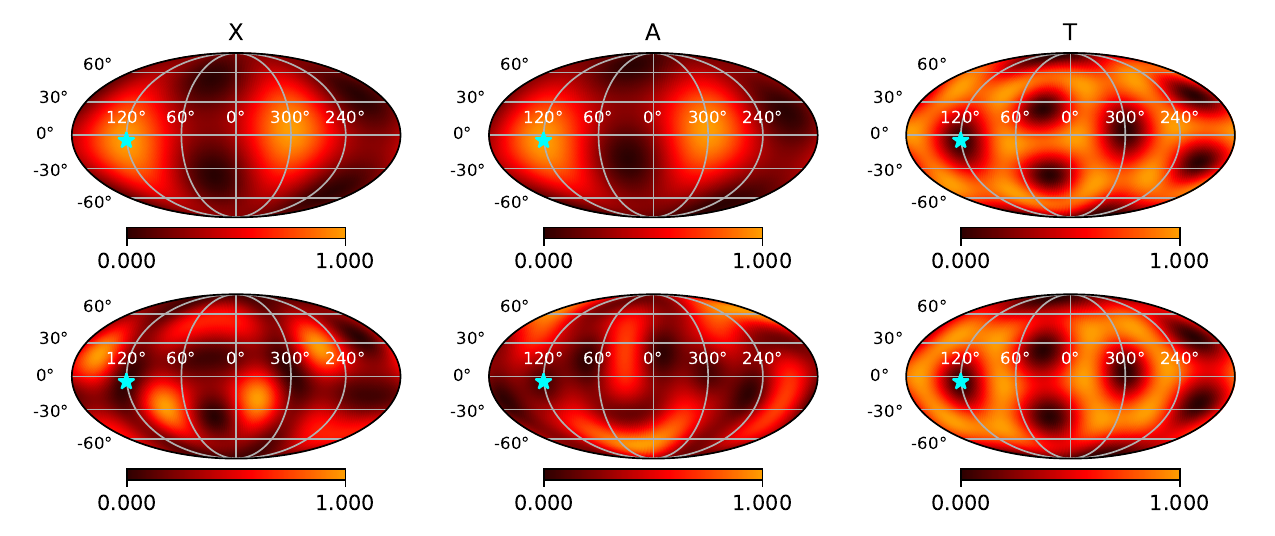}
	\caption{Antenna pattern of different channels for TianQin, which is plotted on a Mollweide projection of the sky in ecliptic coordinates. In this representation, the blue star signifies the pointing direction of TianQin ([lon, lat]$=[120.5\degree,-4.7\degree]$). Top and bottom panels correspond to frequencies of $f^{\rm TQ}_{*}/3$ and $3f^{\rm TQ}_{*}$, respectively.}
	\label{TQ_AET_channels}
\end{figure*}

In~\fig{TQ_AT_ORF}, we present the \ac{ORF} for individual channels of TianQin. 
Given the symmetry inherent in the channel set, the \acp{ORF} for the X/Y/Z channels and the A/E channels are identical, respectively. 
For the detection channels X/Y/Z and A/E, the \ac{ORF} scales with $f^{2}$, while the \ac{ORF} of the null channel T scales with $f^{8}$. 
At lower frequencies, the \ac{ORF} of the null channel is markedly inferior compared to that of the detection channels. 
Nonetheless, as the frequency increases, the \ac{ORF} of the null channel can dramatically rise, thereby transitioning into a detection role. 
\fig{TQ_AET_channels} further illustrates the corresponding antenna pattern when TianQin is positioned at perihelion, within the ecliptic coordinates~\cite{Hu:2018yqb}. 
For clarity, the antenna pattern is standardized relative to its apex, termed the hot spot. 
At the low-frequency threshold ($f=f^{\rm TQ}_{*}/3$), the hot spots of the antenna pattern for detection channels converge with the fixed orientation of TianQin, designated by the coordinates (lon, lat)$=(120.5\degree,-4.7\degree)$. 
As the frequency ascends to $3f^{\rm TQ}_{*}$, the hot spots migrate towards the lateral directions where TianQin is oriented. For the null channel, the antenna pattern demonstrates its minimal intensity in the direction TianQin is facing. 
Moreover, the overall intensity distribution across the celestial sphere remains relatively consistent despite the frequency surge.

\section{Map-making analysis}\label{sec:method}
This section is bifurcated into two pivotal aspects: the initial part explores the utilization of the maximum likelihood method to ascertain the optimal estimator for the \ac{SGWB}'s dirty map, while the subsequent part delves into the application of \ac{SVD} to transform a dirty map into a clean map.

\subsection{Maximum likelihood method}\label{maps}
The data $d_{I}$ is the linear addition of the \ac{SGWB} signal $h_{I}$ and the noise $n_{I}$. 
In the frequency domain, we have
\begin{equation}
    \widetilde{d}_I(f,\tau) = \widetilde{h}_{I}(f,\tau) + \widetilde{n}_I(f,\tau).
\end{equation}
The correlation measurement can be constructed by correlating two data:
\begin{equation}
	\label{eq:cr_mm}
    D_{IJ}(f,\tau) = \frac{2}{\tau}d_I(f,\tau)d^*_J(f,\tau),
\end{equation}
The factor of 2 is consistent with the definition of one-sided \ac{PSD}. 
Combined with Eqs.~(\ref{eq:hh}),~(\ref{eq:Ph}), and~(\ref{eq:hf}), the expectation of the correlation measurement is given by~\cite{Tsukada:2022nsu,Liang:2022ufy}:
\begin{equation}
	\label{eq:D_IJ}
	\langle D_{IJ}(f,\tau)\rangle
	=\bar{H}(f)\int {\rm d}^2\,\hat{\Omega}_{\hat{\mathbf{n}}}\gamma_{I J}(f,\tau, \hat{\mathbf{n}})\mathcal{P}_{\rm h}(\hat{\mathbf{n}})+N_{IJ}(f,\tau).
\end{equation}
To ensure the expectation value attributes pure signals, one can further define the measurement as follows~\cite{Liang:2024ulf}:
\begin{equation}
	C_{IJ}(f,\tau) = D_{IJ}(f,\tau)-N_{IJ}(f,\tau).
\end{equation}

In this work, we utilize the pixel basis grounded in HEALPix ~\cite{Gorski:2004by}. 
HEALPix, a grid-based methodology, is employed for the analysis and visualization of celestial data. 
It is uniquely adept at managing spherical data, characterized by its capacity to uniformly segment the sphere into equal-area and iso-latitude regions. 
Within this paradigm, the angular distribution
\begin{equation}
	\begin{aligned}
		\mathcal{P}_{\rm h}(\hat{\mathbf{n}})=\mathcal{P}_{\hat{\mathbf{n}}'}\delta(\hat{\mathbf{n}},\hat{\mathbf{n}}').
	\end{aligned}
\end{equation}
Then, the expectation of and variance of the above measurement $C_{IJ}$~\cite{Liang:2021bde}
\begin{equation}
	\begin{aligned}
		\label{eq:mu_IJ}
		\mu_{IJ}(f,\tau)=&\langle C_{IJ}(f,\tau)\rangle\\
		=&\sum_{\hat{\mathbf{n}}}\bar{H}(f)\gamma_{IJ}(f,\tau, \hat{\mathbf{n}})
		\mathcal{P}_{\hat{\mathbf{n}}}\\
		\sigma^{2}_{IJ}(f,\tau)
		=&
		\langle C^{2}_{IJ}(f,\tau) \rangle-\langle C_{IJ}(f,\tau) \rangle^{2}\\
		=&
		\frac{1+\delta_{IJ}}{2\,T\Delta f}N_{II}(f,\tau)N_{JJ}(f,\tau)W_{IJ}(f,\tau)
	\end{aligned},
\end{equation}
where $\Delta f$ denotes the frequency resolution, the correction function $W_{IJ}$ is specifically designed for scenarios involving a strong \ac{SGWB}:
\begin{equation}
	\begin{aligned}
		W_{IJ}(f,\tau)
		&=1+\frac{\mu_{II}(f,\tau)N_{JJ}(f,\tau)+\mu_{JJ}(f,\tau)N_{II}(f,\tau)}{N_{II}(f,\tau)N_{JJ}(f,\tau)}\\
		&+
		\frac{\mu_{II}(f,\tau)\mu_{JJ}(f,\tau)+
		(1-\delta_{IJ})\mu^{2}_{IJ}(f,\tau)}{N_{II}(f,\tau)N_{JJ}(f,\tau)}
	\end{aligned}.
\end{equation}
In terms of~\eq{eq:mu_IJ}, the \ac{SNR} is expressed as follows~\cite{Liang:2023fdf}:
\begin{equation}
	\label{eq:snr_tot}
	\rho_{IJ}=
	\sqrt{\sum_{f,\tau}
	\frac{\mu^{2}_{IJ}(f,\tau)}
	{\sigma^{2}_{IJ}(f,\tau)}},
\end{equation}
and the log-likelihood for the $\mathcal{P}_{\hat{\mathbf{n}}}$ is given by
\begin{equation}
	\label{eq:llh_f}
	\ln \left[\mathcal{L}(\mathbf{n})\right]={\rm const.}
	-\sum_{f\tau}\frac{\left|C_{IJ}(f,\tau)-\bar{H}(f)\gamma_{IJ}(f,\tau, \hat{\mathbf{n}})
	\mathcal{P}_{\hat{\mathbf{n}}}\right|^{2}}{2\,\sigma^{2}_{IJ}(f,\tau)}.
\end{equation}

The process of maximizing the likelihood function is fundamentally aligned with optimizing the \ac{SNR}. 
Consequently, the estimator that maximizes the likelihood for the clean map $\mathcal{P}_{\hat{\mathbf{n}}}$ can be derived as~\cite{Thrane:2009fp}:
\begin{equation}
	\label{eq:Pa}
	\hat{\mathcal{P}}_{\hat{\mathbf{n}}}
	=(F^{-1})^{IJ}_{\hat{\mathbf{n}}\hat{\mathbf{n}}'}X^{IJ}_{\hat{\mathbf{n}}'},
\end{equation}
where the dirty map represents the \ac{GW} sky as observed by the detectors:
\begin{equation}
	\label{eq:X_IJ}
	X^{IJ}_{\hat{\mathbf{n}}'}=
	\sum_{f\tau}\gamma^{*}_{IJ}(f,\tau,\hat{\mathbf{n}}')
	\frac{\bar{H}(f)}{\sigma^{2}_{IJ}(f,\tau)}C_{IJ}(f,\tau),
\end{equation}
and the covariance matrix of the dirty map
\begin{equation}
	F^{IJ}_{\hat{\mathbf{n}}\hat{\mathbf{n}}'}
	=\sum_{f\tau}
	\gamma_{IJ}(f,\tau,\hat{\mathbf{n}})\frac{\bar{H}^{2}(f)}{\sigma^{2}_{IJ}(f,\tau)}
	\gamma^{*}_{IJ}(f,\tau,\hat{\mathbf{n}}').
\end{equation}
It is crucial to note that the inverse of $F^{IJ}_{\hat{\mathbf{n}}\hat{\mathbf{n}}'}$ serves as the covariance matrix for the clean map:
\begin{equation}
	\langle \hat{\mathcal{P}}_{\hat{\mathbf{n}}}\hat{\mathcal{P}}^{*}_{\hat{\mathbf{n}}'}\rangle-
	\langle \hat{\mathcal{P}}_{\hat{\mathbf{n}}} \rangle
	\langle \hat{\mathcal{P}}^{*}_{\hat{\mathbf{n}}'} \rangle
	\approx
	(F^{-1})^{IJ}_{\hat{\mathbf{n}}\hat{\mathbf{n}}'}.
\end{equation}
Hence, $F^{IJ}_{\hat{\mathbf{n}}\hat{\mathbf{n}}'}$ is commonly referred to as the \ac{FIM}. 
Furthermore, when considering multiple channel pairs $\{IJ\}$, the aggregation of these individual contributions leads to the formation of the total dirty map and total \ac{FIM}:
\begin{equation}
	\begin{aligned}
		X_{\hat{\mathbf{n}}}^{\rm tot}
		&=&\sum_{IJ}X_{\hat{\mathbf{n}}}^{IJ}\\
		F^{\rm tot}_{\hat{\mathbf{n}}\hat{\mathbf{n}}'}
		&=&\sum_{IJ}F^{IJ}_{\hat{\mathbf{n}}\hat{\mathbf{n}}'}
	\end{aligned}.
\end{equation}

\subsection{Deconvolution}\label{Deconvolution and regularization}
\eq{eq:Pa} indicates the possibility of recovering the clean map by inverting the \ac{FIM} $F^{IJ}_{\hat{\mathbf{n}}\hat{\mathbf{n}}'}$. 
However, the \ac{FIM} is often a singular matrix, necessitating regularization for inversion. 
A common method to invert the \ac{FIM} involves a \ac{SVD} regularization scheme. 
Given the Hermitian nature of the \ac{FIM}, its \ac{SVD} representation takes the form
\begin{equation}
	F=U\Sigma U^{*},
\end{equation}
where $U$ is a unitary matrix, and $\Sigma$ is a diagonal matrix containing positive real eigenvalues $s_{i}$ of the \ac{FIM}. 
When arranging the diagonal elements of $\Sigma$ in descending order, a threshold $s_{\rm min}$ can be selected to condition the matrix. 
Values below this threshold can be typically replaced with infinity or the smallest eigenvalue above the cutoff, resulting in the matrix $\Sigma'$. 
Following this procedure, one can directly derive the regularized $F'$ using the resulting matrix $\Sigma'$:
\begin{equation}
	F'=U\Sigma' U^{*},
\end{equation}
with its inverse
\begin{equation}
	F'^{-1}=U\Sigma'^{-1}U^{*}.
\end{equation}
By multiplying the inverted-regularized \ac{FIM} with the dirty map, the clean map can be derived:
\begin{equation}
	\label{eq:pfx}
	\hat{\mathcal{P}}_{\hat{\mathbf{n}}}=
	(F'^{-1})^{\rm tot}_{\hat{\mathbf{n}}\hat{\mathbf{n}}'}X^{\rm tot}_{\hat{\mathbf{n}}'}.
\end{equation}
The process of discarding eigenvalues during \ac{SVD} can lead to the loss of information associated with pixels of weak intensity, which can result in the generated clean map differing from the true sky map. 
In addition, when solving the equation represented by~\eq{eq:pfx}, round-off errors are introduced. 
These errors are inherently related to the condition number of the \ac{FIM}.

Within the framework of a system of equations denoted as $Ax = b$, the condition number $\kappa(A)$ associated with the coefficient matrix $A$ functions as an indicator of singularity, illustrating its susceptibility to perturbations in the input data. 
It is defined as follows~\cite{Agarwal:2021gvz,kincaid2009numerical}:
\begin{equation}
	\label{eq:cn}
	\kappa(A) = \|A\|\|A^{-1}\|,
\end{equation}
where $\|...\|$ denotes the norm of matrix. 
In numerous applications, the 2-norm is often preferred due to its direct relationship with the eigenvalues of the matrix:
\begin{equation}\label{2-norm}
	\|A\|_{2} = \sqrt{\lambda_{\rm {max}}(A^*A)},
\end{equation}
where $\lambda_{\rm{max}}$ labels the maximum eigenvalue of the matrix. 
Given that the \ac{FIM} usually exhibits off-diagonal elements with negligible imaginary parts, it can be accurately approximated as a symmetric, positive definite matrix. 
Under this assumption, the condition number $\kappa(A)$ of the matrix simplifies to the ratio of its maximum to minimum eigenvalues:
\begin{equation}
	\label{eq:cne}
	\kappa(A) = \lambda_{\rm {max}}(A)/\lambda_{\rm {min}}(A).
\end{equation}
The condition number measures the extent to which the resulting clean map $\hat{\mathcal{P}}_{\hat{\mathbf{n}}}$ fluctuates in response to changes in the dirty map $X$. 
A high condition number suggests that the clean map is highly sensitive to even slight alterations in the dirty map, making it difficult to obtain accurate results. 
Conversely, a low condition number implies that changes in the dirty map have a minimal impact on the clean map, indicating that the \ac{FIM} is ``well-behaved" and thus more effective at recovering the clean map.

\section{Set up and results}\label{sec:result}
In this section, we will demonstrate the capability of TianQin in recovering the sky map for the \ac{SGWB}. 
We narrow our attention to the pixel-based decomposition, with a specific emphasis on point sources. 

\begin{figure}
	\centering
	\includegraphics[width=0.48\textwidth]{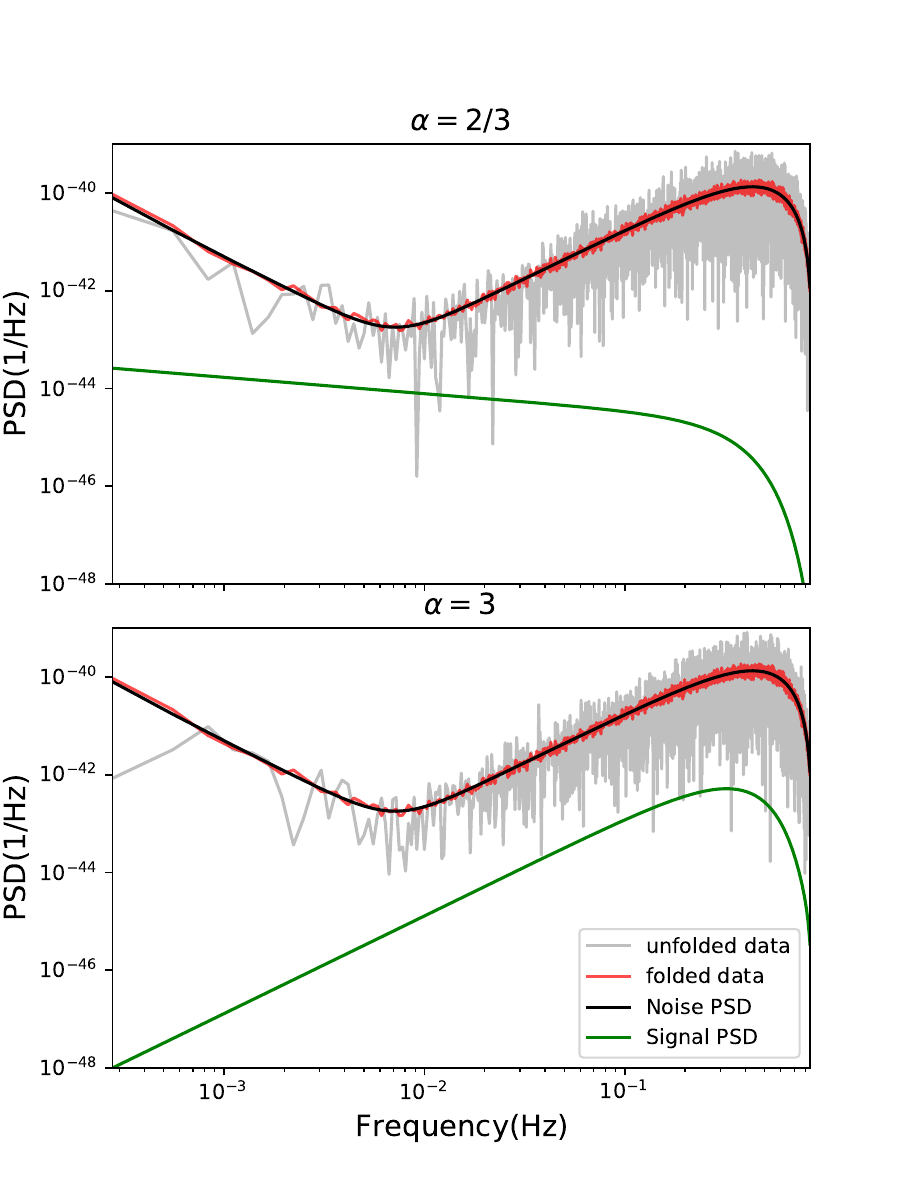}
	\caption{Auto-correlation data for the X channel of TianQin, with the top and bottom panels involving spectral indices $\alpha$ for \ac{SGWB} at 2/3 and 3. Within each panel, the unfolded data is symbolized by the gray line, whereas the folded data is illustrated by the red line. Additionally, the black and green lines denote the noise and signal \acp{PSD}, respectively, corresponding to an \ac{SNR} of approximately 16 over the course of 1-year operation.}
	\label{fig:d_tq}
\end{figure}
\subsection{Data simulation}
We start by generating random Gaussian \ac{SGWB} signals, of which the covariance is obtained by summing contributions from each pixel across the all sky:
\begin{equation}
    S_{IJ}^{\rm in}(f,\tau) = \frac{4\pi}{N_{\rm pix}}\sum_{\hat{\mathbf{n}}}\gamma_{IJ}(f,\tau, \hat{\mathbf{n}})\bar{H}^{\rm in}(f)\mathcal{P}^{\rm in}_{\hat{\mathbf{n}}},
\end{equation}
where the intensity of \ac{SGWB} is determined by the spectral shape $\bar{H}^{\rm in}(f)$ and the angular distribution$\mathcal{P}^{\rm in}_{\hat{\mathbf{n}}}$. 
$N_{\rm pix}$ denotes the number of pixels. 
Increasing the $N_{\rm pix}$ enhances resolution but also escalates computational demands. 
To strike a balance, we opt for $N_{\rm pix}^{\rm in} = 768$ for both injecting and subsequently outputting the signal. 
In addition, to handle the computational load associated with analyzing the real-time response of the detector, which is subject to temporal variations, we divide the data into discrete time segments. 
We assume that within each brief interval, the detector's response remains consistent. 
To encapsulate the full segment's response, we employ the response at the midpoint of the interval as a transient representation. 
For TianQin, we select segments of 3600 seconds, where the directional change in response is minimal~\cite{Liang:2024tgn}. 
This allows a compact period of 3.64 days to be divided into 87 segments. 
On the other hand, we simulate the random Gaussian frequency-domain noise based on the noise \ac{PSD} provided in~\eq{eq:N_xyz}. 
By combining the generated signal with the noise, we derive the correlation measurement $D_{IJ}(f)$ as described in~\eq{eq:cr_mm}.

Next, we examine data gathered during the operational cycle of TianQin, which entails a three-month on phase followed by a subsequent three-month off phase. 
Incorporating a year's worth of data thus yields 50 periods, and these 50 periods can be then aggregated into a single period for analysis. 
By aggregating multiple periods into a single unit, one can significantly reduce computational effort. 
While this method may result in the loss of some low-frequency data, it is deemed acceptable as the data lies outside the sensitive frequency range of TianQin. 
The aforementioned {\it data folding}~\cite{Ain:2015lea,Ain:2018zvo} process is equivalent to rearranging the time summation in~\eq{eq:X_IJ}, maintaining the formulaic expression without alteration. 
\fig{fig:d_tq} presents a sample of the data from the auto-correlated X channel for \ac{SGWB} signal with different spectral indices $\alpha$. 
The gray line represents the unfolded data from a single period, while the red line showcases the folded data after averaging across the 50 periods. 
For comparison purposes, we also plot the \acp{PSD} of the noise and \ac{SGWB} signal using black and green lines, respectively. The \acp{SNR} for the signals with $\alpha$ values of 2/3 and 3 are approximately 16, achieved over the course of TianQin's 1-year operation. 

\begin{figure*}
	\centering
	\includegraphics[width=\textwidth]{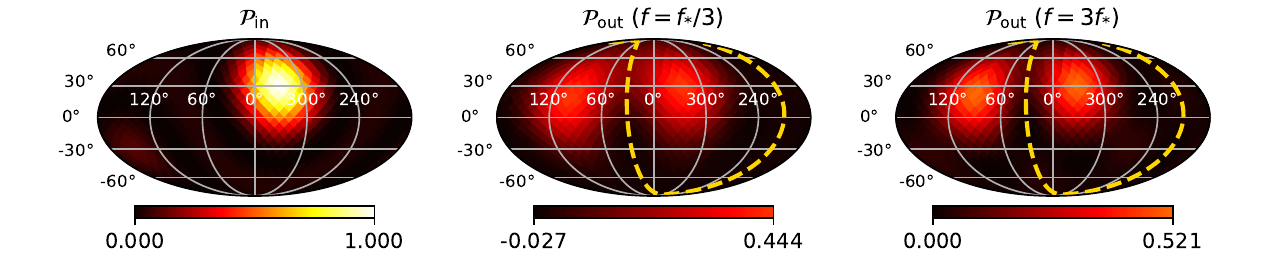}
	\caption{Normalized sky map of a pure \ac{SGWB} signal. The left panel illustrates the injected map for the \ac{SGWB} signal which corresponds to the top panel of~\fig{fig:d_tq}, positioned at [lon, lat]$=[5\pi/3,\pi/6]$. The middle and right panels refer to the recovered map, with the data truncated at $f^{\rm TQ}_{*}/3$ and $3f^{\rm TQ}_{*}$, respectively. The orange dashed line indicates the orbital plane of TianQin, serving as a visual reference to highlight the symmetry observed in the recovered maps.}
	\label{fig:TQ_XYZ_S}
\end{figure*}

\begin{figure*}
	\centering
	\includegraphics[width=0.80\textwidth]{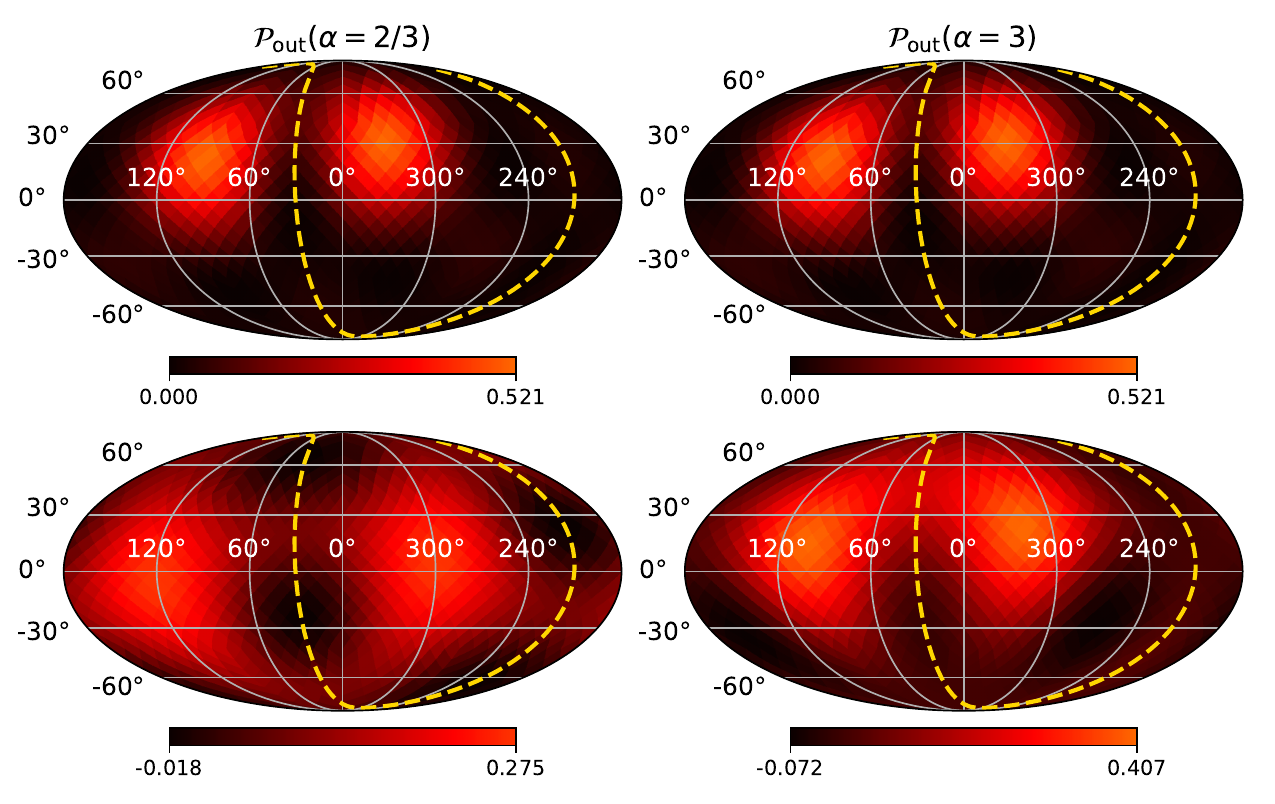}
	\caption{Recovered sky map for the injected \ac{SGWB} signal, as referenced in~\fig{fig:d_tq}. The top and bottom panels correspond to the pure-signal and signal-plus-noise scenarios, respectively. In both cases, the cutoff frequency of the data is set to $3f^{\rm TQ}_{*}$. The orange dashed line signifies the orbit plane of TianQin.}
	\label{fig:TQ_XYZ_SN}
\end{figure*}

\subsection{Map-making}
The process now pivots towards map-making. 
Unless otherwise specified, we will employ a Gaussian, stationary, unpolarized point source expanded up to multipole moments of $l$ up to 4, with the signal \ac{PSD} illustrated in~\fig{fig:d_tq}. 
As depicted in~\fig{fig:TQ_XYZ_S}, we initiate our analysis by injecting a pure signal with $\alpha=2/3$ into the XYZ channel set for data simulations. 
The left panel displays the resulting map after injection, with \ac{SGWB} signal located at [lon, lat]$=[5\pi/3,\pi/6]$. 
It is noteworthy that, to ensure a non-negative distribution of spherical harmonics, Clebsch-Gordan coefficients are employed~\cite{Banagiri:2021ovv,ARFKEN2013773}. 
The middle and right panels exhibit the clean maps recovered by truncating the data to $f^{\rm TQ}_{*}/3$ and $3f^{\rm TQ}_{*}$, respectively. 
For straightforward comparison, these maps have been normalized to the brightest point of the injected map.

The fixed orbital orientation of TianQin renders it incapable of distinguishing between two signals that are symmetrically aligned with the orbital plane. 
As a result, upon injecting a point signal, two point signals are symmetrically recovered relative to the orbital plane (marked by the orange dashed lines), with the intensity of these recovered signals approximating half of the original injected signal. 
This mirror symmetry poses a substantial hurdle in precisely ascertaining the true orientation of the signal. 
We will further explore this issue in Sec.~\ref{sec:Doppler} by analyzing it through the lens of the Doppler effect.

Furthermore, a noteworthy trend in the antenna pattern $\gamma_{IJ}$ manifests as the increasing prominence of the imaginary component with rising frequency. 
This evolution is notably significant as it amplifies the process of inverting the \ac{FIM}, a pivotal procedure for capturing fine-scale details\footnote{Further insights will be expounded upon in~Sec.~\ref{sec:sma}.}. 
As a result, determining the optimal frequency truncation becomes imperative. 
Under the noise-free premise, as depicted in the middle and right panels of~\fig{fig:TQ_XYZ_S}, the distinction between a cutoff frequency of $f^{\rm TQ}_{*}/3$ and $3f^{\rm TQ}_{*}$ is negligible. 
To rigorously evaluate the effects of frequency truncation, we utilized the \ac{JSD}~\cite{menendez1997jensen}, a metric that quantifies the divergence between two distributions. 
A minimum value for \ac{JSD} of 0 signifies that the two distributions are precisely identical, while a maximum value of $\ln 2$ indicates that the two distributions are entirely distinct. 
With a computed \ac{JSD} value of 0.017 nat, we infer that truncating the frequency to $f^{\rm TQ}_{*}/3$ suffices for noise-free conditions.

The effect of noise on map-making is a substantial concern. 
In~\fig{fig:TQ_XYZ_SN}, we further compare the results of map-making with and without noise in the top and bottom panels, where the data is truncated at $3 f_{*}^{\rm TQ}$. 
For comparative analysis, the left and right panels refer to two different spectral shapes, $\alpha = 2/3$ and $\alpha = 3$, respectively. 
In the presence of noise, the capability to map the \ac{SGWB} with $\alpha = 2/3$ is significantly hindered, resulting in an increased \ac{JSD} of 0.127 nat. 
Conversely, when $\alpha = 3$, the recovered map becomes more accurate, with a \ac{JSD} of 0.062 nat. 
This trend aligns with the observations in~\fig{fig:TQ_XYZ_S}, where an increase in the cutoff frequency led to the gradual absorption of the high-frequency components of the signal. 
The transition from $\alpha = 2/3$ to $\alpha = 3$ noticeably intensifies the high-frequency components, which in turn facilitates the results for map-making. 

\begin{figure*}
	\centering
	\includegraphics[width=0.95\textwidth]{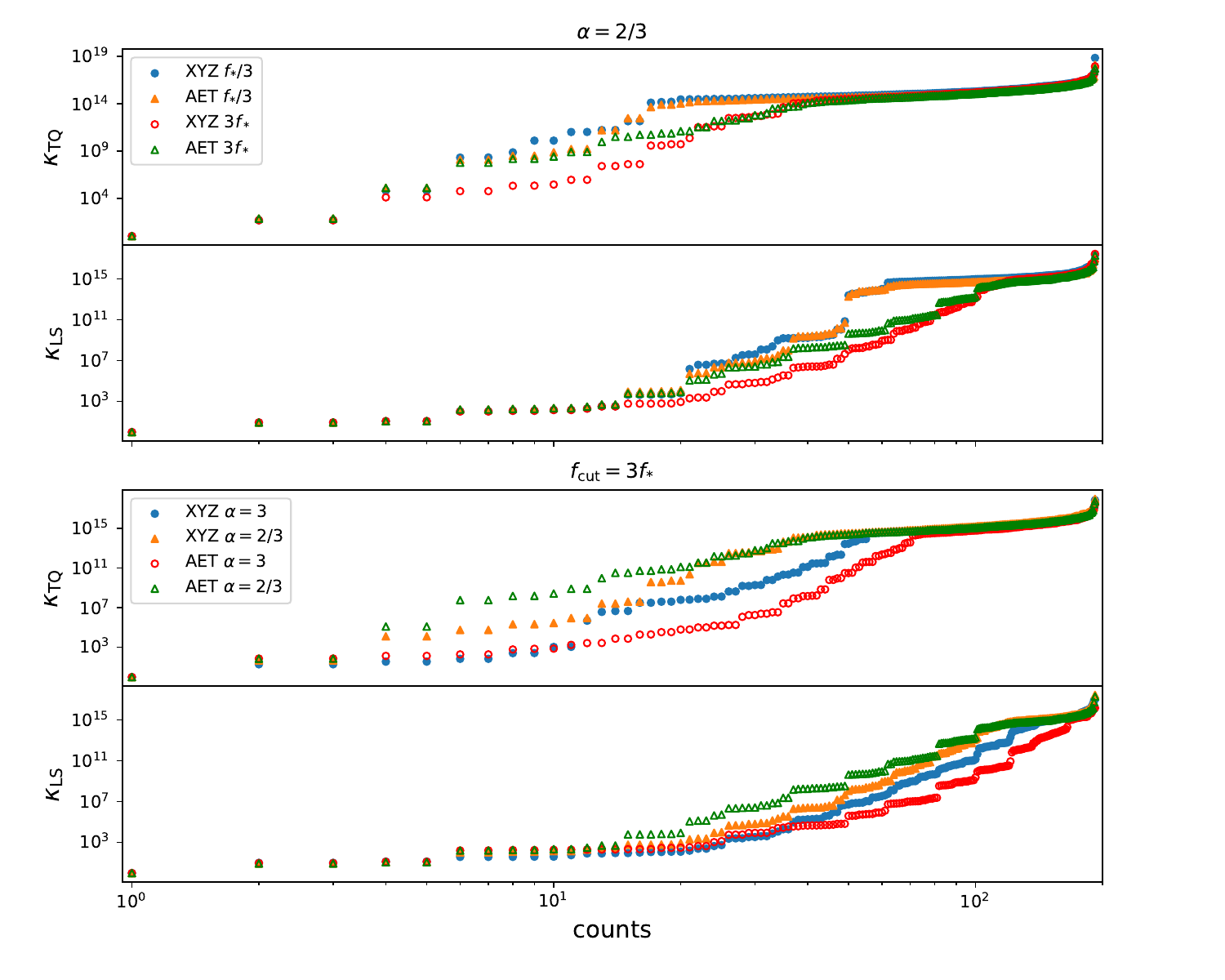}
	\caption{Condition number of \ac{FIM} for different channel sets of TianQin and \ac{LISA}. In the top panel, 
	we display the condition number while holding the spectral index $\alpha$ constant, investigating the influence of altering the cutoff frequency and selecting different channel sets. The bottom panel maintains a constant cutoff frequency at $3f_{*}$, which allows us to examine the impact of different spectral indices on the condition number.}
	\label{Fig:TQ_LS_FN}
\end{figure*}
\subsection{Doppler-induced effects}\label{sec:Doppler}
As previously discussed, TianQin faces a challenge due to mirror symmetry in map-making, as shown in~\fig{fig:TQ_XYZ_S} and~\fig{fig:TQ_XYZ_SN}. 
\eq{eq:Omega_D} implies that Doppler shifts can serve as a useful complementary tool for investigating the frequency profile of the \ac{SGWB} and its inherent rest-frame anisotropies. 
We will explore strategies to tackle this symmetry issue in the context of Doppler effects.

Coupled with the Earth's velocity of $3\times 10^4\,\,\rm m/s$, the space-borne detector's motion towards the signal direction leads to a frequency increase in the real signal and a decrease in the frequency of the symmetric signal. 
This variation in frequency is instrumental in determining the precise location of a point source. 
Considering the point source shown in~\fig{fig:TQ_XYZ_S}, we calculate the Doppler shift using~\eq{eq:dpl_D}. 
Our results indicate that the Doppler effect can cause frequency alterations and directional shifts of approximately $6 \times 10^{-5}$ and $2\times 10^{-5}$, respectively. 
However, when comparing the top and bottom panels of~\fig{fig:TQ_XYZ_SN}, it becomes evident that noise-induced errors reach a level of $10^{-1}$. 
Consequently, we conclude that the effects caused by the Doppler shift are not strong enough to overcome the mirror symmetry for the map-making process.

\subsection{Singular matrix analysis}\label{sec:sma}
The condition number of the \ac{FIM} plays a critical role in the map-making process. 
Given that variations in the \ac{FIM} can occur across different channels, it is essential to conduct a comparative analysis. 
To this end, we plan to utilize both the XYZ and AET channel sets for this study. 
Our goal is to evaluate and understand the differences in their performance and effectiveness within the map-making process.

In~\fig{Fig:TQ_LS_FN}, we calculate the condition number of the \ac{FIM} using~\eq{eq:cne}. 
The eigenvalues of the matrix are plotted in descending order, and we vary the number of eigenvalues retained for comparative analysis. 
It is important to recognize that while retaining a higher count of eigenvalues incorporates more information, it generally inflate the condition number of the \ac{FIM}. 
Therefore, our analysis focuses on the condition numbers associated with the first 192 eigenvalues.

The top panel compares the condition numbers for TianQin and LISA at cutoff frequencies of $f_*/3$ and $3f_*$, with a fixed spectral index $\alpha$ of 2/3. 
The condition number for \ac{LISA} consistently remains lower than that for TianQin, especially at higher eigenvalue counts. 
The discrepancy can be attributed to the differing operational designs of the detectors: LISA is capable of surveying various directions in the sky throughout the year, offering dynamic coverage that enhances its condition number performance. 
In contrast, TianQin's operational focus remains fixed on J0806, leading to a stable and less varying antenna pattern. 
This stability significantly inflates the condition number of the \ac{FIM}. 
Despite these differences, both detectors share certain characteristics in the map-making process. 
As expected, there is a notable correlation between an increase in the cutoff frequency and a decrease in the condition number, regardless of whether the XYZ or AET channel set is used.  Interestingly, when the cutoff frequency is set at $f_*/3$, the XYZ channel set typically exhibits a higher condition number compared to the AET channel set. 
However, the result dynamic changes when the cutoff frequency is elevated to $3f_*$, with the AET channel set then showing a higher condition number. 
The selection of both cutoff frequency and channel set plays a crucial role in influencing the condition number of the \ac{FIM} for map-making.

The results presented above are predicated on a specific spectral index, $\alpha=2/3$. 
Should this index vary, the outcomes would necessarily adjust to accommodate such changes.  
To illustrate this point, the bottom panel keeps the cutoff frequency constant at $3f_*$, and compares the results for $\alpha=2/3$ with those for $\alpha=3$. 
A distinct pattern is observed: with $\alpha=2/3$, both TianQin and LISA tend to show a higher condition number for the XYZ channel set compared to the AET channel set; 
however, the situation is reversed when $\alpha = 3$. 
This observation underscores that the spectral shape of the \ac{SGWB} has a significant impact on the accuracy and reliability of the map-making process.

\section{Summary}\label{sec:summary}
In this paper, we have explored the map-making for the anisotropic \ac{SGWB} using TianQin. 
We utilized the maximum likelihood method to construct maps from simulated data. 
By using a point source as a case study, we reconstructed its clean map up to multipole moments of order $l\le4$ both in the absence and presence of noise, achieving an \ac{SNR} of 16. 
We observed that the inclusion of noise necessitates the use of higher-frequency data to achieve satisfactory map-making results. 
A notable limitation of TianQin lies in distinguishing between a point source and its mirror position relative to the orbital plane. 
To tackle this challenge, we investigated the potential of harnessing Doppler-induced effects to break this symmetric degeneracy. 
However, the theoretical deviation in the signal between two symmetrical directions, induced by the Doppler effect, is considerably smaller than the statistical error introduced by the noise. 
Distinguishing signals originating from symmetrical directions remains a tough challenge.

Regarding both TianQin and \ac{LISA}, we further investigated various factors that could influence the map-making process by analyzing the condition number of the \ac{FIM} with respect to three key aspects: the cutoff frequency of the data, the choice of channel set, and the spectral profile of the \ac{SGWB}. 
Our results indicated that adopting a sufficiently high cutoff frequency, specifically at $3f_{*}$, can improve the mapping capabilities. 
Notably, at this cutoff frequency, the XYZ channel set can outperform the AET channel set in terms of map-making efficacy when the spectral index $\alpha$ is set at 2/3. 
Additionally, the map-making process for an \ac{SGWB} with a spectral index $\alpha$ of 3 consistently yielded better results compared to an index $\alpha$ of 2/3.

It is noteworthy that the injection of a point source represents an idealized scenario, selected because it provides a straightforward example to analyze the detection capabilities for anisotropic \ac{SGWB}. 
Furthermore, we focused on single space-borne detectors, underscoring the critical necessity for accurate noise level estimation to facilitate effective \ac{SGWB} detection. 
Failure to do so can significantly undermine the accuracy of detection~\cite{Muratore:2023gxh}. 
A strategy to address this issue entails cross-correlating data from multiple detectors~\cite{Hu:2023nfv,Liang:2024tgn}. 
Unlike ground-based detectors, the complex relative motion of space-borne detectors substantially amplifies the computational demands for determining antenna patterns during the map-making process. 
Research into this topic is ongoing and will be expounded upon in forthcoming publications.

\begin{acknowledgments}
This work has been supported by the National Key Research and Development Program of China (No. 2020YFC2201400), the Natural Science Foundation of China (Grants No. 12173104), the Natural Science Foundation of Guangdong Province of China (Grant No. 2022A1515011862), and the Guangdong Basic and Applied Basic Research Foundation(Grant No. 2023A1515030116). 
Z.C.L. is supported by the Guangdong Basic and Applied Basic Research Foundation (Grant No. 2023A1515111184). 
\end{acknowledgments}

\normalem
\bibliographystyle{apsrev4-1}
%%%%%%%%%%%%%%%%%%%%%%%%%%%%%%%%%%%%%%%%%%%%%%%%%%%%%%%%%%%%%%%%
%%% ½«²Î¿¼ÎÄÏ×Ìí¼Óµ½ reference.bib ÎÄ¼þÀï£¬ÔÚÕâÀïµ÷ÓÃ %%%%%%
%%%%%%%%%%%%%%%%%%%%%%%%%%%%%%%%%%%%%%%%%%%%%%%%%%%%%%%%%%%%%%%%
\bibliography{ref}
\end{document}